# We Drink Good 4.5-Billion-Year-Old Water

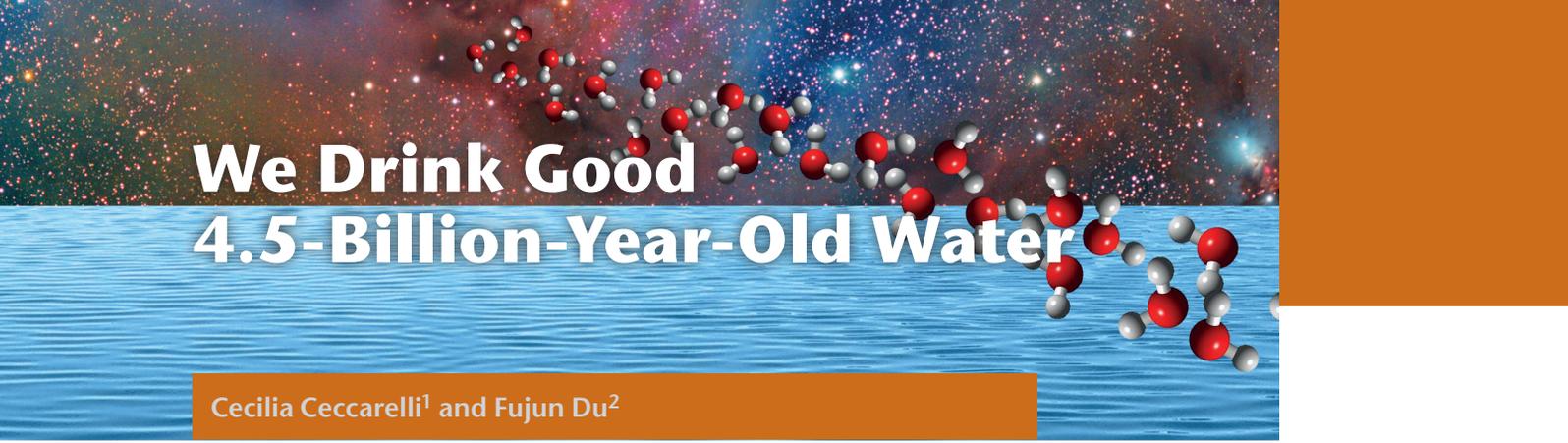

Cecilia Ceccarelli[1] and Fujun Du[2]

**W**ater is crucial for the emergence and evolution of life on Earth. Recent studies of the water content in early forming planetary systems similar to our own show that water is an abundant and ubiquitous molecule, initially synthesized on the surfaces of tiny interstellar dust grains by the hydrogenation of frozen oxygen. Water then enters a cycle of sublimation/freezing throughout the successive phases of planetary system formation, namely, hot corinos and protoplanetary disks, eventually to be incorporated into planets, asteroids, and comets. The amount of heavy water measured on Earth and in early forming planetary systems suggests that a substantial fraction of terrestrial water was inherited from the very first phases of the Solar System formation and is 4.5 billion years old.

## INTRODUCTION

The Solar System was born 4.5 billion years ago, after a long and complex process that led to the planets, asteroids, and comets we see today. Reconstructing what happened is not simple—after all, we only have at hand the final products of the process. It is like deciphering the recipe of a tasty complex cake from the single piece of it that we eat. Only an experienced cook might succeed, exactly because of their experience with other cakes. For astronomers, it is the same: we can try to reconstruct the history of the Solar System by gaining experience with other planetary systems presently forming in the Milky Way.

The study of star and planet formation is one of the most recent fields in astronomy. A major reason is that early forming planetary systems are cold objects that emit photons in the millimeter to infrared wavelengths, which are largely absorbed by the terrestrial atmosphere. Although many aspects of Solar System formation are still poorly understood, the major phases of the process are now well established (e.g., Caselli and Ceccarelli 2012). Most relevant for this article, the history of water in the Solar System is intrinsically connected to these stages, which can be summarized in four major steps, as illustrated in FIGURE 1.

**STEP 1**: The story starts in a cold (~10 K) and relatively dense (~$10^5$ molecules/cm$^3$) clump of a molecular cloud in the Milky Way, the seed from which a solar-type planetary system is born (Caselli et al. 2012). As in average galactic clouds, gas (whose most abundant element is hydrogen, followed by helium, oxygen, and carbon) is mixed with tiny (~0.1 μm in radius) grains of silicate and carbonaceous dust. Given the very low temperatures, when oxygen atoms or molecules encounter the dust grains, they remain frozen on the surface. On the contrary, the lighter H atoms hop around the grain surfaces, and when they encounter the frozen oxygen, they react and form water (e.g., Dulieu et al. 2010), as schematically shown in FIGURE 2. In this first step, almost all available elemental oxygen becomes trapped into water ice.

**STEP 2**: Under the gravitational force, the molecular cloud clump becomes increasingly dense until matter freely collapses towards the center, forming a so-called protostar. The gravitational energy is now transformed into heat and, at a few tens of astronomical units (au: 1 au is the average distance of the Earth from the Sun) from the protostar center, dust is heated up to a temperature around 100 K, where the frozen water enveloping the grains sublimates. In these regions, called hot corinos, whose sizes are similar to that of the Solar System, water becomes the most abundant molecule (Ceccarelli 2004). To provide an order-of-magnitude sense of the water abundance, a typical hot corino contains about 10,000 times the water in the Earth's oceans. So, at this stage, there is plenty of water, even though it is gaseous and not liquid (see also van Dishoeck et al. 2021).

**STEP 3**: The initial small rotation of the molecular cloud clump eventually causes the formation of a disk of material around the nascent star, called a protoplanetary disk. This is a crucial phase because the water that formed in STEP 1 and released into the gas in STEP 2 largely recondenses in the coldest zones of the disk. There, the dust grains become again enveloped by icy mantles, in which the previous history is now stored. Thus, dust grains are the guardians of water inheritance.

**STEP 4**: Planets, asteroids, and comets form in protoplanetary disks from the agglomeration of the dust grains. Some of them contain the oldest water-ice with the imprint of its origin, as we will see later in this article. The water trapped in the agglomerated grains is subsequently released in the rocky planets and kept frozen in distant and cold asteroids and comets.

1 Université Grenoble Alpes CNRS, IPAG
  F-38000 Grenoble, France
  E-mail: Cecilia.Ceccarelli@univ-grenoble-alpes.fr

2 Purple Mountain Observatory
  Chinese Academy of Sciences
  10 Yuanhua Road, Qixia District
  Nanjing 210023, China
  E-mail: fjdu@pmo.ac.cn

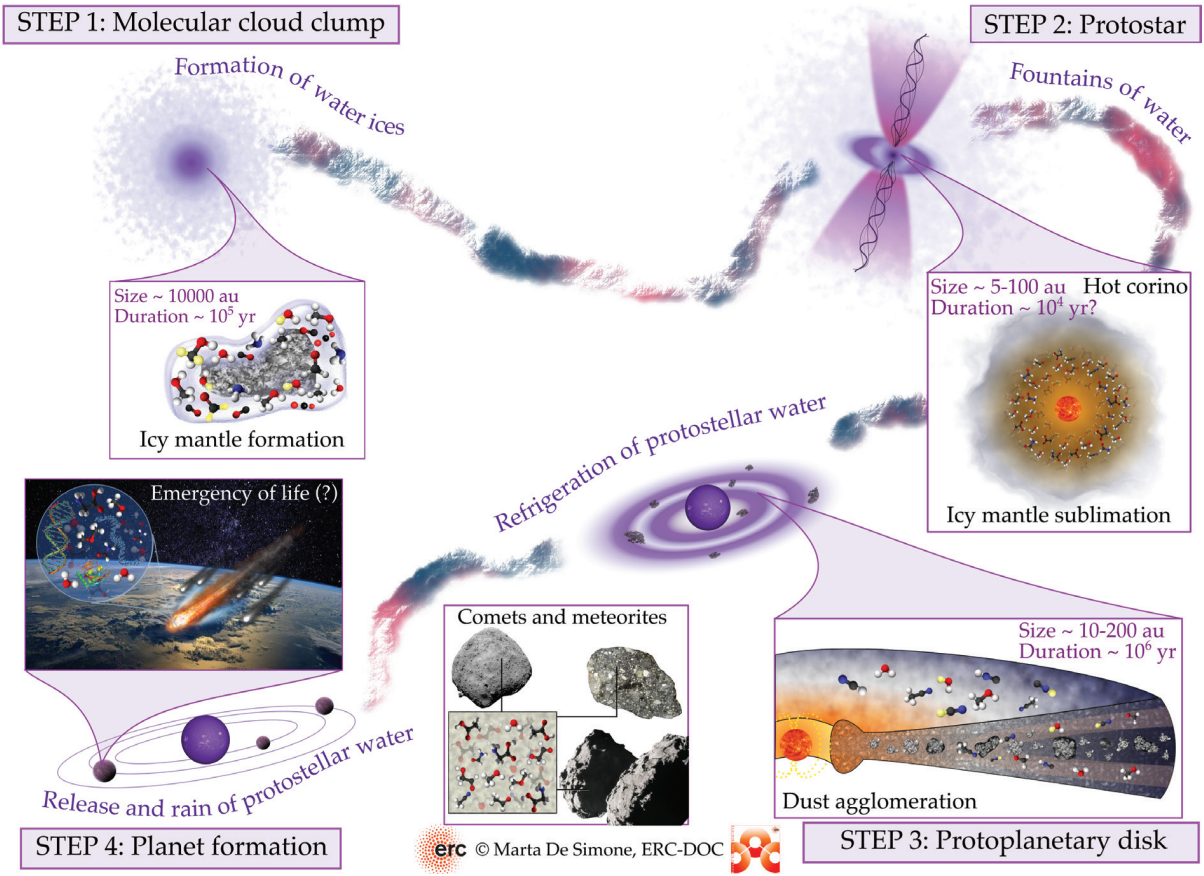

**Figure 1** Sketch of the four major steps involved in the formation of a planetary system similar to the Solar System and of water (Courtesy of M. De Simone), as described in the text. STEP 1: In cold (≤10 K) and dense clumps, water is formed on the surface of the interstellar grains. STEP 2: In the hot corinos of protostars, water is observed in a quantity equivalent to more than 10,000 terrestrial oceans. STEP 3: In protoplanetary disks, the hot corino water is frozen into ices enveloping the grains. STEP 4: The dust grains coagulate into larger rocks that eventually become planets, asteroids, and comets; some of the ices fabricated in STEPs 1 and 2 are transmitted in this way to the final planetary system.

## INTERSTELLAR FACTORIES OF HEAVY WATER

As briefly explained above, most of the water in the protostellar and protoplanetary phases is synthesized during the cold phase of STEP 1 (Fig. 1). This process, illustrated in Figure 2, is relatively simple: hydrogen atoms landing on the grain surfaces hydrogenate oxygen by successive additions. Yet, the extreme conditions of the matter in the molecular cloud clumps, caused by the very low temperature (~10 K), trigger a very peculiar phenomenon, called super-deuteration (Ceccarelli et al. 2014).

Deuterium, the heavy isotope of hydrogen, is (only) produced during the first seconds after the Big Bang, in a ratio of about 1/100,000 with respect to H. If it was statistically distributed in water, the abundance of heavy water (HDO), would be then about $10^{-5}$ times that of normal water ($H_2O$). However, in hot corinos, the $HDO/H_2O$ ratio is only a bit less than 1/100. To make things even more extreme, the doubly deuterated water $D_2O$ is 1/1000 with respect to $H_2O$, namely about $10^7$ times larger than what would be estimated from the D/H elemental abundance ratio (Coutens et al. 2012).

Basically, these extremely large deuterated ratios are caused by the enhanced number of D atoms with respect to H atoms landing on the grain surfaces at the moment of water formation. In turn, the root of the enhancement is the asymmetry of the reaction $H_3^+ + HD \rightarrow H_2D^+ + H_2$ with respect to its reverse reaction $H_2D^+ + H_2 \rightarrow H_3^+ + HD$. The latter is inhibited at low (<20–30 K) temperatures because of the lower zero-point energy of $H_2D^+$ with respect to $H_3^+$; thus, deuterium atoms are transferred from HD, the major D reservoir in cold molecular gas, to $H_2D^+$, then to D, and eventually to water (e.g., Tielens 1983). There are no other ways to obtain this large amount of heavy water in hot corinos nor in general. Therefore, abundant heavy water is a hallmark of water synthesis in the cold molecular cloud clump during the STEP 1 era.

## PROTOPLANETARY DISKS, WHERE PLANETS ARE BORN

During the early stages of star formation (spanning STEPS 3 to 4), young stars are surrounded by a protoplanetary disk of gas and dust, in which planets are forming. Their existence is due to the conservation of angular momentum (Hartmann 2009). The original molecular cloud in which stars form inevitably has a certain amount of angular momentum. As the cloud shrinks, it spins, maintaining a centrifugal force, which impedes radial mass flow so that matter tends to infall parallel to the rotation axis, and in this way a flattened disk is formed. Vertically, the disk is supported by thermal pressure. The mass of a disk is usually about two orders of magnitude lower than that of the central star; hence, it is the stellar gravity that dominates the rotation dynamics of the disk. In such nearly Keplerian rotation, the angular velocity is a decreasing function of radius. The angular velocity gradient leads to internal friction in the disk, which causes angular momentum and matter to transfer. Gas and dust gradually migrate and are either accreted into the central star, assimilated into the

## INTERSTELLAR WATER FORMATION

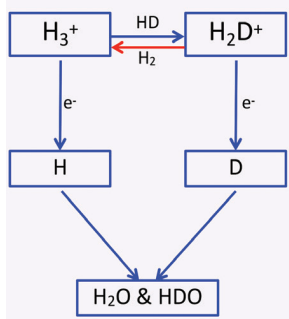
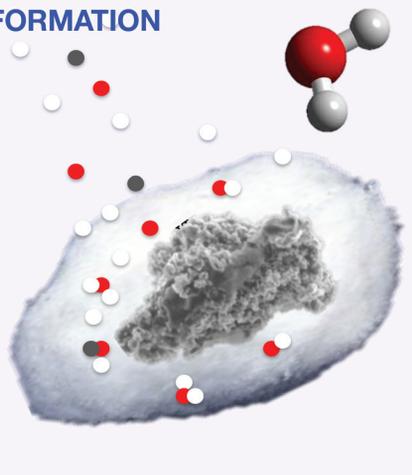

**FIGURE 2** Scheme of interstellar normal ($H_2O$) and heavy water (HDO) formation on the surface of dust grains. **(LEFT)** The HDO abundance surpasses that of the available D/H ratio because of the reaction between $H_3^+$ and HD, the reservoir of deuterium, whose reverse reaction is inhibited at low (<20–30 K) temperatures (red arrow). **(RIGHT)** Dust grains, whose refractory cores have a radius of about 0.1 μm, end up enveloped by icy mantles with a thickness of approximately 0.05 μm, enriched in heavy water. Atoms and molecules are not drawn to scale in this figure.

forming planets, or may escape to outer space. During these processes, delicate structures can be formed, some of which are directly imaged by state-of-the-art telescopes, such as the Atacama Large Millimeter/submillimeter Array (ALMA) (e.g., Andrews et al. 2018). The first image showing rings and gas in the dust distribution of protoplanetary disks is shown in FIGURE 3. Finally, the disk dissipates and a planetary system is born. The protoplanetary disk stage usually lasts a few million years.

As a remnent of the star formation process, the initial chemical composition of the disk largely inherits that of the original molecular cloud clump (STEP 1). Carbon and oxygen easily combine to form carbon monoxide molecules, with the remaining oxygen mostly going into water, either in the gas phase or ice. It is tempting to think that water in protoplanetary disks will end up in planets that are being made. However, the actual history of terrestrial water may not be so simple (see also Izidoro and Piani 2022 this issue).

Protoplanetary disks are heated mainly by the central young star (FIG. 4). Energetic photons (from visible to ultraviolet light and X-ray) emitted by the star can be scattered and/or absorbed by the dust particles in the disk. Dust particles get heated when gaining energy and re-emit photons. The re-emitted photons tend to be at longer wavelengths than the absorbed ones, because the dust grains cannot be as hot as the central star. As these re-emitted photons propagate outwards, they are repeatedly absorbed and scattered, becoming less energetic and eventually escaping from the disk. The disk is thus hotter in the inner region and colder in the outer region. The exact thermal structure depends on many aspects. For example, heating by turbulent viscosity within the disk and by external UV and cosmic rays may also play a role, and the total mass and distribution of matter in the disk also influence how photons propagate in the disk, which determines the thermal structure.

Deriving a disk's thermal structure is generally a complicated task and is calculated using sophisticated computer codes (e.g., Oka et al. 2011). However, in a very simplified model, where the disk is assumed to be optically thin (i.e., nearly transparent for light propagation), the disk temperature can be easily calculated by balancing the projected stellar flux with the re-emitted flux at each location of the disk. For the present-day Solar System, the temperature calculated this way is 278 K at the radial distance of the Earth and 170 K at 2.7 AU (Hayashi 1981). This sets the "snow line" of water in the present-day Solar System. When the Solar System was younger, however, its disk was cooler and the snow line was closer to the center, up to ~1 AU (Morbidelli et al. 2016). Water mainly exists as gas inside the snow line, and is largely, but not entirely, frozen beyond the snow line. Different molecules condense at different temperatures, hence they have different snow lines. Besides being a chemical division boundary, the snow lines may also have dynamical effects. For example, it has been found that the gap locations in the HL Tau disk coincide with the condensation front of ammonia and clathrate hydrates (Zhang et al. 2015). One explanation for this coincidence is that, at the snow lines, dust grains coagulate more efficiently, creating the observed patterns of rings and gaps. The concept of a snow line is related to the "habitable zone" of a planetary system, in the sense that the latter is usually defined based on the existence of liquid water. However, the origin and sustenance of life depends on many factors (e.g., stable magnetic field, protection against frequent asteroid impact, tectonics) and a definition based merely on the existence of liquid water may not be satisfactory.

## STEAMING HOT WATER VAPOR

It is well established that a large amount of warm (>100 K) water vapor exists in the inner region of protoplanetary disks (FIG. 5). While the ground-based observation of astronomical water is usually hindered by the water vapor in the Earth's atmosphere, the inner region of protoplanetary disks could be considerably warmer than the Earth's atmosphere, which can excite transitions that are not significantly absorbed by the atmosphere.

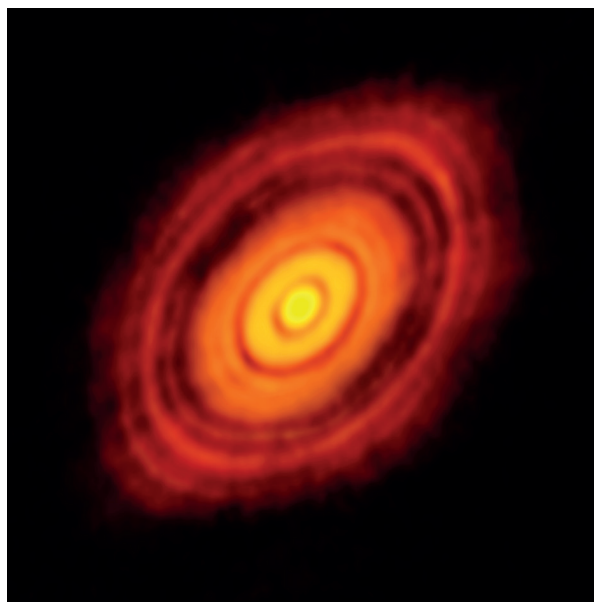

**FIGURE 3** This image from the Atacama Large Millimeter/submillimeter Array (ALMA) shows fine details in a planet-forming disk surrounding a young star, HL Tau. The rings and gaps in the disk are thought to be related to forming planets.

The observed water vapor lines are fitted with sophisticated models to obtain a temperature of the emitting region of a few hundred to one thousand Kelvin (Salyk et al. 2008; Pontoppidan et al. 2010). The excitation conditions and Doppler line broadening of the water lines show that they are emitted from a region spanning a few au around the central star. The spectral feature of some water lines shows that part of the water vapor may be in disk winds (Salyk et al. 2019). The two hydrogen atoms of water molecules generally have parallel or anti-parallel spins. This leads to the existence of two forms of water molecules, ortho and para, which is a purely quantum mechanical effect. It is difficult to change the spin orientation of hydrogen atoms at low (<100 K) temperature, thus the relative abundance of ortho and para water only depends on the temperature at which the water formed. Lower temperatures are associated with smaller ortho-to-para ratios, thus its measure from water lines can be used to infer at which temperature the water formed, namely its formation temperature.

disk, becomes chemically inactive because the dust grains covered by water ice coagulate and settle down into the midplane of the disk, thus depleting the atmosphere of the outer disk of oxygen. These detailed processes have been modeled (e.g., Krijt et al. 2020). The shortage of oxygen also leads to a reduction of carbon monoxide, which is indeed seen in many disks (Long et al. 2017). As a consequence of oxygen depletion, carbon, which may also be depleted but not as severely, is set free from the carbon monoxide molecule and combines with hydrogen through a chain of reactions to form hydrocarbons. This is seen in ALMA[1] images as hydrocarbon rings (Bergin et al. 2016).

Could there be "wet" dust grains (i.e., grains covered with liquid water) in some area between the hot and cold regions of a protoplanetary disk? A glance at the phase diagram of water informs that this would be very unlikely. At a temperature of ~280 K, the minimum vapor pressure for water to be in the liquid phase is 1 kPa, which amounts to a water vapor number density of approximately $3 \times 10^{17}$ molecules/cm$^{-3}$ and a corresponding total gas density (mostly hydrogen) that would still be at least four orders of magnitude higher; a density rarely seen in protoplanetary disks.

## HEAVY WATER AND ARIADNE'S THREAD

As shown throughout this article, water is synthesized in large quantities during the early phases of the formation of a planetary system similar to our own and, likewise, at the birth of the Solar System. A crucial question is whether and how much of this 4.5-billion-year-old water reached Earth.

There are two major observational quantities that help to constrain theories on the origin of terrestrial water: the amount of water and its degree of deuteration, namely, the ratio of heavy over normal water, HDO/$H_2O$. In principle, more than enough water formed during STEPS 1–3 and the question is rather how much it could have been transmitted to Earth (see Izidoro and Piani, 2022, this issue). Here we focus on the second observational parameter, water deuteration, which provides an incredibly strong constraint on terrestrial water. FIGURE 6 provides an instructive hint as

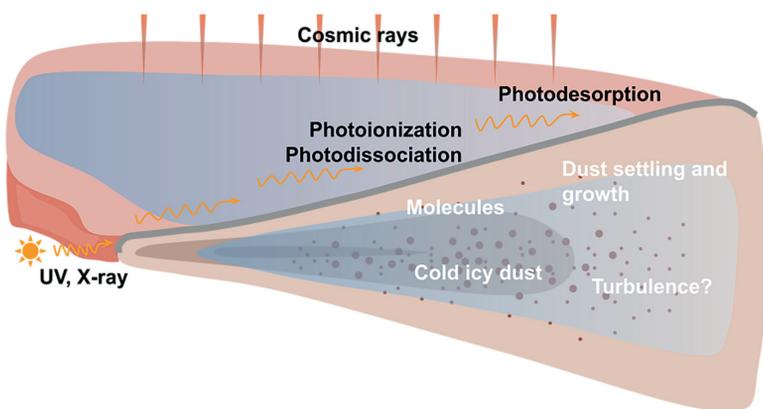

**FIGURE 4** Schematic view of a protoplanetary disk. External cosmic rays and photons from the central star are shown as straight and curly arrows, respectively. The distribution of molecules and dust grains is shown, and the existence of turbulence is also marked.

## FREEZING COLD WATER

Water also theoretically exists in the gas phase in the outer part of a disk, where the temperature is extremely cold, (e.g., Dominik et al. 2005; Woitke et al. 2009; Du and Bergin 2014), although its abundance (relative to hydrogen) can be very low (~$10^{-10}$–$10^{-9}$). The temperature of dust grains is 10–30 K in these regions; thus, the frozen water is not able to thermally sublimate. However, cosmic rays may play a crucial role. While cosmic rays may be attenuated by the magnetic field of the star (Cleeves et al. 2014), they could still deposit a significant amount of energy to the dust grains, which drives water into the gas phase. The low density of the outer region also allows the scattered UV photons from the star to penetrate, which helps desorb water.

Unfortunately, the theoretical predictions described above do not quite match the observations. For example, one deep search with the Herschel Space Observatory showed that cold water vapor emission is only firmly detected in 2 out of 13 disks (Du et al. 2017), with only upper limits for the other 11. Even in the two positively detected cases, the measured values are much lower than model predictions. To explain this discrepancy, after exploration of the parameter space, the total abundance of oxygen available to chemical processes has to be reduced by up to two orders of magnitude relative to normal astrochemical models. The rationale behind this is that the element oxygen, presumably mostly in the form of water ice in the outer to what probably happened by comparing the HDO/$H_2O$ values in terrestrial water with those of hot corinos—the only objects where HDO has so far been detected in early forming, solar-type planetary systems—as well as in Solar System objects, namely comets, Enceladus, and meteorites (Ceccarelli et al. 2014). "Heavy over normal" water on Earth is about 10 times larger than the elemental D/H ratio in the Universe and consequently at the birth of the Solar System, in what is called the solar nebula. As explained above, this is a very precise indication that at least part of the water that arrived to us formed in the original molecular cloud clump from which the Solar System was born, as observed in hot corinos. Based on Figure 6, approximately 1% to 50% of terrestrial water was inherited from the very first phases of the Solar System's birth (see also Cleeves et al. 2014).

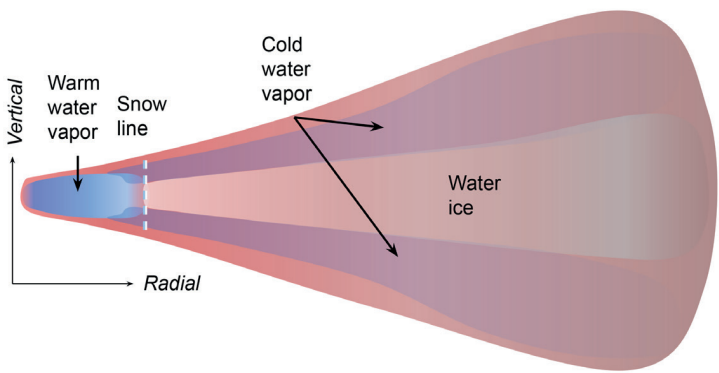

**FIGURE 5** Schematic cross-sectional view showing the distribution of different phases of water in a protoplanetary disk. Warm water vapor, with temperatures higher than ~170 K, exists in the inner disk inside the snow line, and cold water vapor, whose temperature is at most 30 K, and ice coexist in the outer disk.

---

1 ALMA (Atacama Large Millimeter/submillimeter Array) is a radio interferometer composed of 66 antennas located in the Atacama desert of Chile.

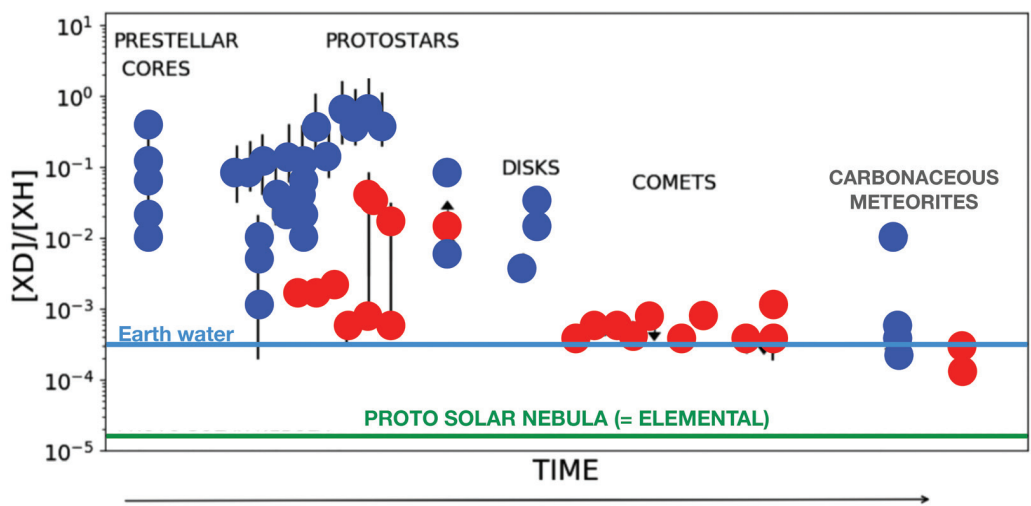

**Figure 6** Molecular deuteration in early forming planetary systems and Solar System objects (adapted from Bianchi et al. 2019). The water deuteration, given by the HDO/$H_2O$ ratio (red symbols), diminishes from the hot corino stage to the comets and finally to the asteroid whose fragments populate the carbonaceous meteorites. The latter have exactly the same HDO/$H_2O$ ratio as the terrestrial value (light blue line) and 10 times the original elemental D/H of the proto-solar nebula (green line) (see Ceccarelli et al. 2014 for more details). Organic material (blue symbols) is also enriched with deuterium because of the same process illustrated in Figure 2. Some of the hydrogen atoms contained within the organic matter could have enriched the final terrestrial water budget (see, e.g., Izidoro and Piani 2022 this issue).

The water in comets and asteroids (from which the vast majority of meteorites originate) was also inherited since the beginning in large quantities (Fig. 6).

Earth likely inherited its original water predominantly from planetesimals, which are supposed to be the precursors of the asteroids and planets that formed the Earth, rather than from the comets that rained on it.

In conclusion, the amount of heavy water on Earth is our Ariadne thread, which can help us to come out from the labyrinth of all possible routes that the Solar System may have taken.

## OLD GOOD WATER

While isotopic evidence (namely, the HDO/$H_2O$ ratio) suggests that Earth's water has its origin 4.5 billion years ago and was probably brought to Earth with planetesimals, the exact processes of how this happens remain far from clear. The issue is quite involved because the origin and evolution of Earth's water is inevitably connected with other important participants on this planet, e.g., carbon (e.g., Mikhail and Füri 2019), molecular oxygen, and the magnetic field. They are all part of the same history, from the origin of worlds to the origin of life. The dynamical and chemical roles played by water (and other volatile species) before and during planet formation also require clarification, namely, how important they are in aiding planetesimal growth, and how they help sequester other molecules by forming ice mantles or clathrates.

Here we presented a simplified early history of the Earth's water according to the most recent observations and theories. A good fraction of terrestrial water likely formed at the very beginning of the Solar System's birth, when it was a cold cloud of gas and dust, frozen and conserved during the various steps that led to the formation of planets, asteroids, and comets, and was eventually transmitted to the nascent Earth. How the final passage happened is another fascinating chapter, described in Izidoro and Piani (2022 this issue).

## ACKNOWLEDGMENTS

C. Ceccarelli acknowledges funding within the European Union's Horizon 2020 research and innovation programme from the European Research Council (ERC) for the project "The Dawn of Organic Chemistry" (DOC), grant agreement no. 741002, and from the Marie Sklodowska-Curie Fellowship Programme for the project "Astro-Chemical Origins" (ACO), grant agreement no. 811312. Fujun Du is supported by the National Natural Science Foundation of China (NSFC) through grant nos. 11873094 and 12041305.